# A fully flexible joint lattice position and dose optimization method for LATTICE therapy

Xin Tong, Weijie Zhang, Ya-Nan Zhu, Xue Hong, Chao Wang, Jufri Setianegara, Yuting Lin, and Hao Gao

Department of Radiation Oncology, University of Kansas Medical Center, USA

Email: hgao2@kumc.edu

**Acknowledgment**

The authors are very thankful to the valuable comments from reviewers. This research is partially supported by the NIH grants No. R37CA250921, R01CA261964, and a KUCC physicist-scientist recruiting grant.

**Abstract.**

**Background:** Lattice radiotherapy (LATTICE) is a technique of spatially fractionated radiation therapy (SFRT) that delivers high radiation doses to specific regions (vertices) within a large tumor, forming a spatially modulated "lattice" pattern, while surrounding areas receive lower doses to minimize damage to healthy tissues. In LATTICE, vertices (regions of high dose) are typically placed at regular intervals within tumors, such as simple cubic or hexagonal arrangement, which limits the flexibility needed to adapt to irregular tumor shapes and the proximity of critical organs, potentially leading to unexpected hotspots or under-treatment of tumor regions. Optimizing the placement of vertices in LATTICE is beneficial for precisely targeting high-dose regions within the tumor while minimizing radiation exposure to adjacent healthy tissue, but there is still no optimization method available for solving the positions of fully flexible placed vertices. The great challenge in such optimization lies in handling the constraints on the relative positions between different vertices.

**Purpose:** This work aims to develop a new treatment planning method for LATTICE with fully flexible placement of vertices and joint optimization of the position of each lattice vertex and dose, to improve overall plan quality compared with conventional LATTICE planning methods relying on manual regular placements of lattice vertices.

**Methods**: The proposed method jointly optimizes each lattice vertex position and other plan optimization variables (proton spot weights or photon fluences) during the dose optimization process. This is formulated as a new constrained optimization problem by adding each lattice vertex position to optimization variables with appropriate constraints to meet the requirements of the LATTICE vertices placement guideline on the 1) center-to-center distance between lattice vertices and 2) distance of lattice vertices to the target boundaries. The optimization problem is solved by the alternating direction method of multipliers and iterative convex relaxation methods.

**Results:** Plans of our proposed method (NEW) were compared with conventional LATTICE plans. We created 100 LATTICE plans for the abdomen and lung patients, respectively, with different vertices positions, respectively, from which three plans, termed BEST, MID and WORST, were selected for each anatomical location with the largest, median and smallest total optimization objective $F$. All LATTICE plans optimized with the NEW method showed results that comparable to or better plan qualities as BEST. For example, for photon LATTICE abdomen plans, the values of $F$ were 1.92 (NEW), 2.79 (WORST), 2.27 (MID) and 1.96 (BEST) representing a 31.1% improvement from WORST to NEW; the PVDR values were 5.88 (NEW), 3.00 (WORST), 4.33 (MID) and 5.16 (BEST) representing 96.0% and 14.0% improvements from WORST and BEST respectively to NEW.

**Conclusion:** A new LATTICE treatment planning approach with fully flexible joint lattice position and dose optimization is introduced and demonstrated to improve target PVDR and OAR sparing, compared to conventional LATTICE method with regular placements of lattice vertices.

1. Introduction

Spatially fractionated radiation therapy (SFRT) is a treatment modality that intentionally creates spatial dose modulation within the tumor volume and/or organs-at-risk (OAR). Compared to uniform target dose approaches, SFRT has advantages for debulking large tumors for mass effect and pain relief and minimizing radiation-induced side-effects such as nearby OAR toxicities that are associated with uniformly irradiating a large tumor volume [1-12]. Clinically, thousands of patients have been treated with SFRT techniques across multiple anatomical treatment sites with well-established clinical techniques such as GRID and LATTICE [13].

LATTICE is a 3D form of spatial modulation compared to its simpler GRID SFRT counterpart which creates spatial modulation in 2D. These 3D spatial modulations are achieved with multiple beam angles and inverse optimization methods (e.g., IMRT, VMAT or IMPT techniques) using photon or proton beams [14-16]. Typically, LATTICE treatments involve delivering ablative doses (peak) to an array of high-dose vertices (lattice vertices) within the target, while simultaneously delivering a low dose (valley) to the rest of the target. These lattice vertices are typically represented geometrically as spherical target sub-volumes that are spaced apart by a minimum center-to-center (c-t-c) distance to generate a peak-to-valley dose ratio (PVDR) which is clinically desired. For safety purposes, clinical practices involve constraining the positions of these lattice vertices within an inner target sub-volume that is typically 5-10 mm retracted from the target surface.

Currently, LATTICE treatment planning begins with the geometrical design of the lattice array which contains the positional distributions of these lattice vertices with spherical diameters ranging between 0.8 and 1.5 cm and c-t-c distances ranging between 2 and 6 cm. This process is completely manual with the actual design pre-decided and influenced by the target size, adjacent OAR proximities and clinical dose objectives [5,12,14,17,18,19]. It is clinically well-established that the resultant SFRT qualities of the LATTICE plans are highly dependent on the optimal placements (position and geometric arrangements) of these lattice vertices [20-23]. However, the current manual and predetermined fixed placements of these lattice vertices do not consider the actual SFRT dose distributions during dose optimization and are therefore hypothesized to be non-optimal. We thus hypothesize that better SFRT plan qualities are achievable if optimal LATTICE positions are iteratively and simultaneously determined during treatment planning as part of the SFRT dose optimization process.

The purpose of this work is to develop a LATTICE treatment planning method that allows fully flexible high-dose vertices placement and optimizes their positions alongside other plan optimization variables (proton spot weights or photon fluences). In contrast to conventional LATTICE plans, the key feature of our proposed method is that the vertices are not arranged in a regular pattern (cubic or hexagonal) but can be flexibly placed while adhering to the placement constraints in [15]. In a related work [20], although the positioning of the entire lattice grid is optimized, the absolute positions of individual vertices are not optimized, i.e., only rigid movement of the entire lattice grid is allowed. Compared to previous work [20], the novelty and the primary challenge of this work is that many non-convex and complex geometry constraints are introduced into the optimization problem to allow the independent optimization of each vertex position. We will benchmark and compare the proposed full flexible LATTICE optimization method against the conventional regular-pattern LATTICE method for the purpose of improving PVDR and sparing OAR.

## 2. Methods and materials

*2.1. Geometry and parameters*

In this work, lattice vertices are represented as spheres of radius $R$ and are placed within the target. A peak ablative dose $d_{peak}$ is delivered to these spheres while the remaining target receives a valley dose $d_{valley}$. The lattice vertex positions of each sphere are represented by a set of centers $r=\{r_i, 1 \leq i \leq N_r\}$, where $r_i=(r_{ix}, r_{iy}, r_{iz})$ is the center of $i^{th}$ lattice vertex, and $N_r$ is the number of lattice vertices. Then $\{s_j, 1 \leq j \leq N_d, s_j=(s_{jx}, s_{jy}, s_{jz})\}$ is the set of all voxels' coordinates with size number $N_d$. The target dose objective $d$ at the $j^{th}$ voxel is

$$d_j(r) = d(s_j, r) = d_{valley} + (d_{peak} - d_{valley}) \sum_{i=1}^{N_r} I_{\|r_i - s_j\| \leq R}, \quad j = 1, 2, \ldots, N_d, \quad (1)$$

where $\|\cdot\|$ is the Euclidean norm throughout this paper, and $I_{\|r_i - s_j\| \leq R}$ is an indicator function that equals to one if the voxel $j$ is inside a lattice vertex and zero otherwise.

The Eq. (1) is a mathematical definition of the prescription dose, that is, $d_j = d_{peak}$ for the voxel $j$ inside the vertices while $d_j = d_{valley}$ for the voxel $j$ elsewhere. We set $d_{peak} = 15 d_{valley}$ in (1) as the target dose objective, to approximately reach $d_{peak} = 5 d_{valley}$, in the optimized plan.

*2.2. Full-flexible joint lattice position and dose optimization method*

For simplicity, adopting the least squared type of plan objective, the mathematical model of the problem takes the following form

$$\min_{x,r} \|Ax - d(r)\|^2 \quad (2)$$

For photon modalities, $x$ represents the photon fluences via IMRT. For proton modalities, $x$ corresponds to the proton spot weights via IMPT. $A$ is the dose influence matrix for photons or protons. However, the complete form of the plan objective used in this work consists of active-set least squares based on Dose-Volume-Histogram (DVH) constraints [33-36].

To mitigate the differentiability problem of indicator function, replacing the indicator function in (1) by the sigmoid function yields

$$d_j(r) = d_{valley} + (d_{peak} - d_{valley}) \sum_{i=1}^{N_r} \frac{1}{1 + e^{-\lambda(R - \|s_j - r_i\|)}}, j = 1, 2, \ldots, N_d, \quad (3)$$

where $\lambda$ is a controlling parameter that changes the steepness of the sigmoid function. The constraints on $(x, r)$ are

$$x \geq 0, \quad (4a)$$
$$\|r_i - r_j\| \geq L_1, \quad i \neq j, \ i, j \in \{1, 2, \ldots, N_r\}, \quad (4b)$$
$$\text{dist}(r_k, \Gamma) \geq L_2, \quad k = 1, 2, \ldots, N_r, r_k \text{ in target}, \quad (4c)$$

where $L_1$ is the minimum center-to-center distance and $L_2$ is the minimum center to boundary distance. The first inequality enforces the non-negativity of beam weights. Based on current clinical trials [10,12,15], we set the lattice vertex radius $R = 10$ mm. Lattice vertices are at least 10 mm away from each other to prevent intermediate dose bridging, resulting in a minimum center-to-center distance of $L_1 = 2R + 10 = 30$mm. All lattice vertices are constrained within a target sub-volume that is at least 5 mm from the target boundary $\Gamma$ to confer OAR sparing robustness, therefore the distance of a center to boundary is no less than $L_2 = R + 5 = 15$ mm (4c). The structure is displayed in Fig. 1. The initial lattice positions are chosen empirically.

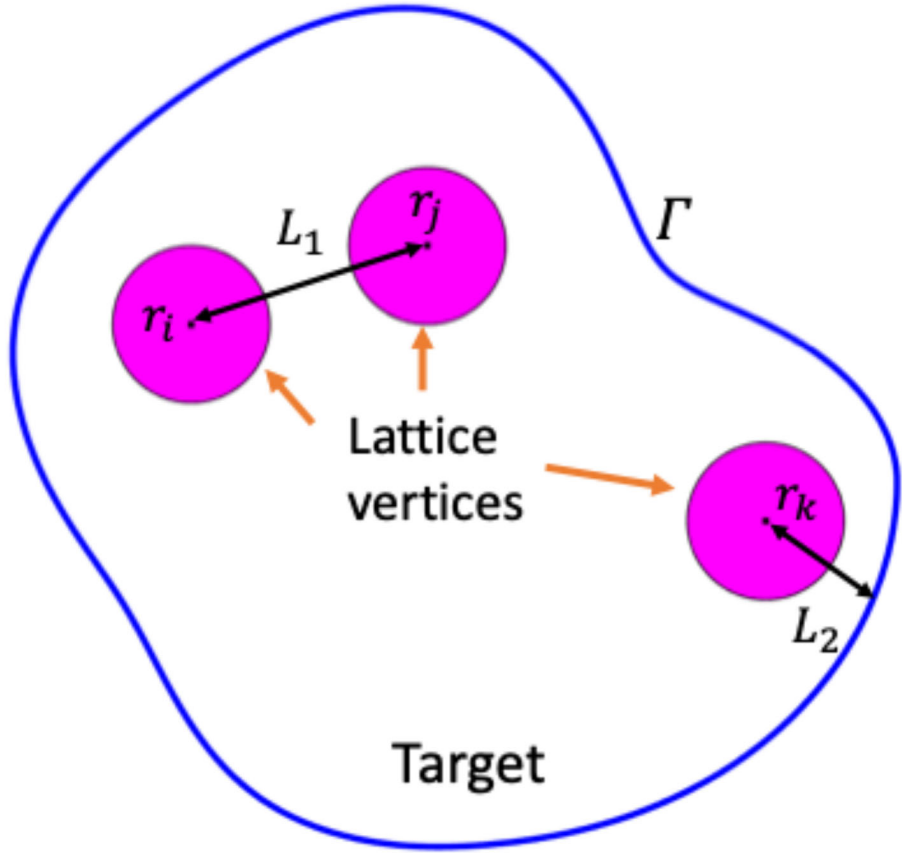

Figure 1. Plot of the cross-section of a target.

With (3) and (4a-4c), the (1) can be rewritten as

$$\min_{x,r} \|Ax - d(r)\|^2$$
$$\text{s.t.,} \begin{cases} x \geq 0, \\ \|r_i - r_j\| \geq 30, \quad i \neq j, \ i,j \in \{1,2,\ldots,N_r\}, \\ \text{dist}(r_i, \Gamma) \geq 15, \quad i = 1,2,\ldots,N_r, r_i \text{ in target.} \end{cases} \quad (5)$$

*2.3. Solution algorithm*

The DVH constraints of Eq. (5) are handled by an iterative convex relaxation method [37-40] and then the alternating direction method of multipliers (ADMM) is used to solve geometric constraints. ADMM is an algorithm designed to solve constrained and convex optimization problems by decomposing the original problem into smaller, more manageable subproblems [24, 25]. It has recently been used to solve various treatment planning problems [20, 26-28].

By introducing the auxiliary variables $y=\{y_{ij}, i{\neq}j, 1{\leq}i{\leq}N_r, 1{\leq}j{\leq}N_r\}$ and $z=\{z_i, 0{\leq}i{\leq}N_r\}$, one can decouple the $x$ and $r$ constraints from the planning objective to their dual variables $u=\{u_{ij}, i{\neq}j, 1{\leq}i{\leq}N_r, 1{\leq}j{\leq}N_r\}$ and $v=\{v_i, 0{\leq}i{\leq}N_r\}$. To define the augmented Lagrangian function for (5) with penalty parameters $\mu_0, \mu_1, \mu_2$, one obtains

$$L(x,r,y,u,z,v) = \|Ax - d(r)\|^2 + \mu_0\|x - z_0 + u_0\|^2$$
$$+ \mu_1\left(\sum_{i\neq j}^{N_r}\|(r_i - r_j) - y_{ij} + u_{ij}\|^2\right)$$
$$+ \mu_2\left(\sum_{i=1}^{N_r}\|r_i - z_i + v_i\|^2\right), \quad (6)$$
$$s.t., \begin{cases} z_0 \geq 0, \\ \|y_{ij}\| \geq 30, & i \neq j, \ i,j \in \{1,2,...,N_r\}, \\ \text{dist}(z_i, \Gamma) \geq 15, & i = 1,2,...,N_r. \end{cases}$$

The constrained optimization problem (5) is solved iteratively by minimizing primal variable $x,r$, and dual variable $z,y,u,v$ of the augmented Lagrangian as follows

$$\begin{cases}
(x^{k+1}, r^{k+1}) = \underset{x,r}{argmin}\, L(x, r, y^k, u^k, z^k, v^k), & (7a) \\
z_0^{k+1} = \underset{z_0}{argmin}\, L(x^{k+1}, r^{k+1}, y^k, u^k, z_0, v_0^k, z_i^k, v_i^k), & (7b) \\
z_i^{k+1} = \underset{z_i}{argmin}\, L(x^{k+1}, r^{k+1}, y^k, u^k, z_0^k, v_0^k, z_i, v_i^k), & (7c) \\
y_{ij}^{k+1} = \underset{y_{ij}}{argmin}\, L(x^{k+1}, r^{k+1}, y_{ij}, u_{ij}^k, z^k, v^k), & (7d) \\
v_0^{k+1} = v_0^k + x^{k+1} - v_0^{k+1}, & (7e) \\
v_i^{k+1} = v_i^k + z_i^{k+1} - v_i^{k+1}, & (7f) \\
u_{ij}^{k+1} = u_{ij}^k + (r_i^{k+1} - r_j^{k+1}) - y_{ij}^{k+1}. & (7g)
\end{cases}$$

While the *(x, r)* subproblem (7a) is nonlinear, it is differentiable with respect to both *x* and *r*. Therefore, it can be effectively solved using the quasi-Newton method [29].

The $z_0$ subproblem (7b) has the following analytic formula,

$$z_0^{k+1} = max(x^{k+1} + u_0^k, 0). \qquad (8)$$

The $z_i$ subproblem (7c) has the following analytic formula,

$$z_i^{k+1} = \begin{cases} r_i^{k+1} + v_i^k, & \text{if dist}(z_i, \Gamma) \geq 15,\ z_i \text{ in target,} \\ \tilde{z}_i, & \text{otherwise.} \end{cases} \qquad (9)$$

where $\tilde{z}_i$ is a projection of $z_i$ onto the target boundary $\Gamma$ that lies 15 mm from $\Gamma$ with $\text{dist}(\tilde{z}_i, \Gamma) = 15$.

The *y* subproblem (7d) has the following analytic formula,

$$y_{ij}^{k+1} = \begin{cases} (r_i^{k+1} - r_j^{k+1}) + u_{ij}^k, & \text{if } \left\|(r_i^{k+1} - r_j^{k+1}) + u_{ij}^k\right\| \geq 30, \\ \dfrac{30 \cdot \left[(r_i^{k+1} - r_j^{k+1}) + u_{ij}^k\right]}{\left\|(r_i^{k+1} - r_j^{k+1}) + u_{ij}^k\right\|}, & \text{otherwise.} \end{cases} \qquad (10)$$

*2.4. Materials*

We evaluated the effectiveness of the proposed approach with fully-flexible lattice ("NEW") by comparing its results against the classical LATTICE approach with regular-pattern lattice. The classical LATTICE plans were generated by exhaustive searching method. Specifically, we optimized the conventional LATTICE treatment planning problem (2) with pre-set lattice arrays $r$. For this evaluation, we selected 100 lattice array sets for an abdomen patient and a lung patient respectively. Next, all solutions were sorted by total optimization objective, from which the plan with the smallest optimization objective ("BEST"), the plan with the median optimization objective ("MID"), and the plan with the largest optimization objective ("WORST") were chosen as the reference SFRT plans to be compared with NEW.

The dose influence matrices $A$ were generated via matRad [30] on 3 mm³ dose grid. Both photon and proton LATTICE therapy were considered. The large bowel is an OAR for the abdomen case, while the esophagus and trachea are OARs for the lung case. For photon LATTICE, the beam angles were angles were from 12º to 348º with an increment of 24º for abdomen and from 0º to 336º with an increment 24º for lung; for proton LATTICE, the beam angles were (60º, 150º, 240º, 330º) for abdomen and (0º, 120º, 240º) for lung. The target dose constraints $d_{valley}$ is 2Gy with $d_{peak}$ = 30Gy. The same plan normalization $d_{mean,valley}$=2Gy is applied to all plans for a fair comparison.

## 3. Results

*3.1. Total plan objective*

The total plan objectives $F$ for abdomen and lung LATTICE plans are presented in Tables 1 and 2, respectively. A significant improvement in $F$ is demonstrated for NEW as compared to WORST and MID for all cases. For example, the $F$ values for the proton LATTICE abdomen plan (proton abdomen) was 1.25 for NEW, compared to 2.20 (WORST), 1.55 (MID) and 1.21 (BEST), demonstrating a 19.4% reduction in median $F$ values among all generated conventional proton abdomen plans when compared to NEW. Across all cases, NEW $F$ values were comparable to or slightly better than BEST.

*3.2. PVDR*

In Table 1, there is a notable increase in PVDR values for NEW as compared to WORST and MID for abdomen patient: from 3.00 (WORST) and 4.33 (MID) to 5.88 (NEW) with photon, from 4.90 (WORST) and 6.91 (MID) to 7.98 (NEW) with proton. When compared to BEST, NEW had comparable PVDR values. A similar result was observed in Table 2 for lung case. The variances in PVDR stemmed from differences in $D_{peak}$, as the $D_{valley}$ values remained consistent across all plans after dose normalization.

Variations in PVDR values can also be observed when comparing axial slices from each plan in Figs. 2-5. For the abdomen case in Fig. 2, NEW exhibited more clearly defined peaks as compared to WORST and MID. One dose peak near the large bowel was diminished in WORST (Fig. 3(a) vs 3(b)) or moved away from the WORST ((Fig. 2(a) vs 2(b)).

*3.3. OAR sparing*

The OARs $D_{mean}$ values as presented in Tables 1-2 demonstrated that, lower OARs $D_{mean}$ for NEW as compared to WORST and MID for most scenarios, with similar OAR $D_{mean}$ when compared to BEST. For example, for the proton lung plan, NEW had smaller esophagus $D_{mean}$ (0.27 Gy) compared to WORST (0.45 Gy), MID (0.43 Gy) and BEST (0.32 Gy).

The enhancements in OAR dose achieved by NEW, in contrast to WORST and MID, were further validated through DVH plots in Fig. 6. The red solid lines (NEW) were lower than the blue dash-dotted lines (WORST) and green dotted lines (MID), and close to the cyan dotted line (BEST).

Table 1. Dosimetric results for the abdomen LATTICE plans, namely: (1) total optimization objective $F$, (2) mean valley dose $D_{valley}$, (3) mean peak dose $D_{peak}$, (4) PVDR defined to be $D_{peak}/D_{valley}$, and (5) mean large bowel dose $D_{bowel}$. Doses are reported in the unit of Gy.

|  |  | NEW | WORST | MID | BEST |
|---|---|---|---|---|---|
| Photon | F | 1.92 | 2.79 | 2.27 | 1.96 |
| Photon | $D_{valley}$ | 2.00 | 2.00 | 2.00 | 2.00 |
| Photon | $D_{peak}$ | 11.77 | 6.01 | 8.65 | 10.32 |
| Photon | PVDR | 5.88 | 3.00 | 4.33 | 5.16 |
| Photon | $D_{bowel}$ | 0.15 | 0.29 | 0.26 | 0.18 |
| Proton | F | 1.25 | 2.20 | 1.55 | 1.21 |
| Proton | $D_{valley}$ | 2.00 | 2.00 | 2.00 | 2.00 |
| Proton | $D_{peak}$ | 15.96 | 9.80 | 13.83 | 16.52 |
| Proton | PVDR | 7.98 | 4.90 | 6.91 | 8.26 |
| Proton | $D_{bowel}$ | 0.07 | 0.13 | 0.10 | 0.06 |

Table 2. Dosimetric results for the lung LATTICE plans, namely: (1) total optimization objective $F$, (2) mean valley dose $D_{valley}$, (3) mean peak dose $D_{peak}$, (4) PVDR defined to be $D_{peak}/D_{valley}$, (5) mean esophagus dose $D_{eso}$ and (6) mean carina doses $D_{carina}$. Doses are reported in the unit of Gy.

|  |  | NEW | WORST | MID | BEST |
|---|---|---|---|---|---|
| Photon | F | 2.15 | 2.29 | 2.20 | 2.12 |
| Photon | $D_{valley}$ | 2.00 | 2.00 | 2.00 | 2.00 |
| Photon | $D_{peak}$ | 9.62 | 8.75 | 9.17 | 9.79 |
| Photon | PVDR | 4.81 | 4.38 | 4.58 | 4.89 |
| Photon | $D_{eso}$ | 0.50 | 0.75 | 0.78 | 0.56 |
| Photon | $D_{carina}$ | 0.56 | 0.74 | 0.52 | 0.55 |
| Proton | F | 1.63 | 1.85 | 1.72 | 1.61 |
| Proton | $D_{valley}$ | 2.00 | 2.00 | 2.00 | 2.00 |
| Proton | $D_{peak}$ | 13.54 | 11.64 | 12.52 | 13.41 |
| Proton | PVDR | 6.77 | 5.82 | 6.26 | 6.70 |
| Proton | $D_{eso}$ | 0.27 | 0.45 | 0.43 | 0.32 |
| Proton | $D_{carina}$ | 0.27 | 0.48 | 0.35 | 0.20 |

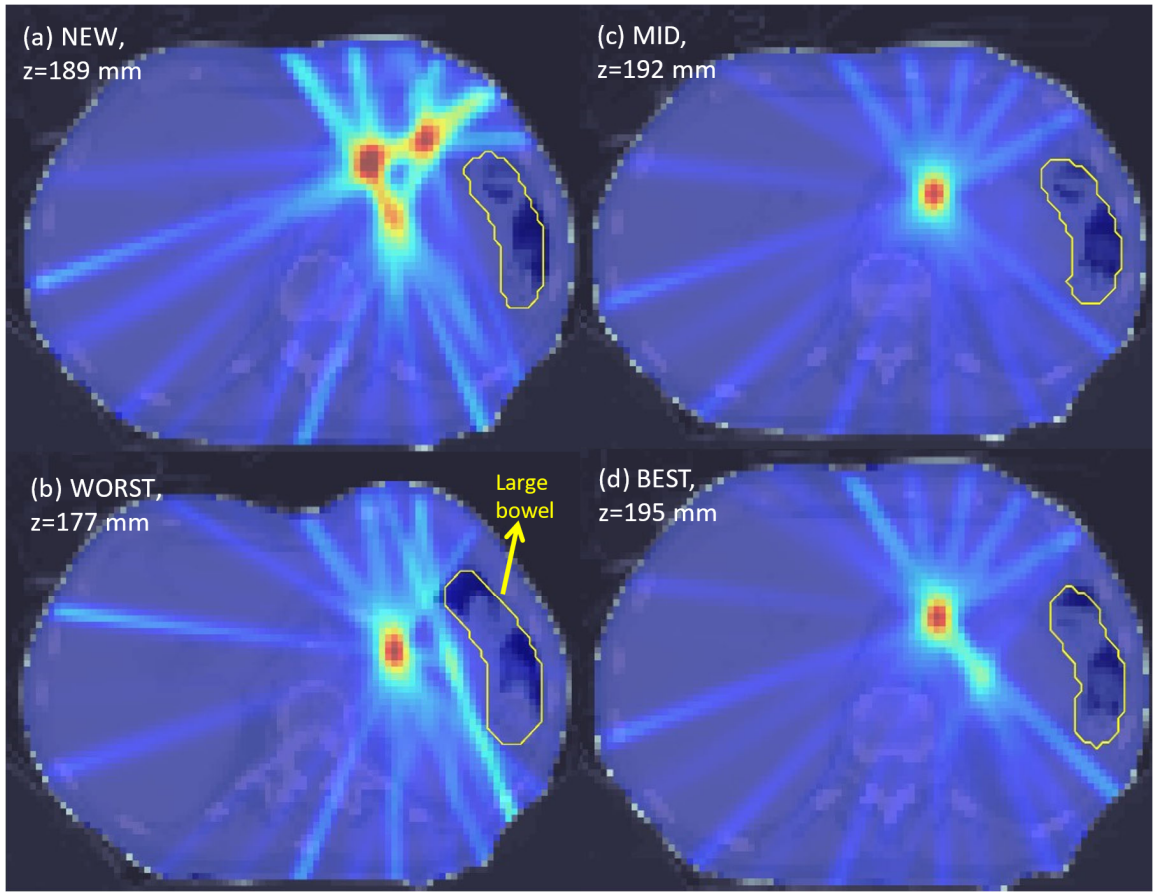

Figure 2. Axial dose slices for photon abdomen LATTICE plans using various techniques namely (a) NEW; (b) WORST; (c) MID; (d) BEST. Due to position differences of the lattice vertices between different techniques, the axial slices of the maximal PVDR from each plan are presented. The dose plot window is [0 Gy, 18 Gy].

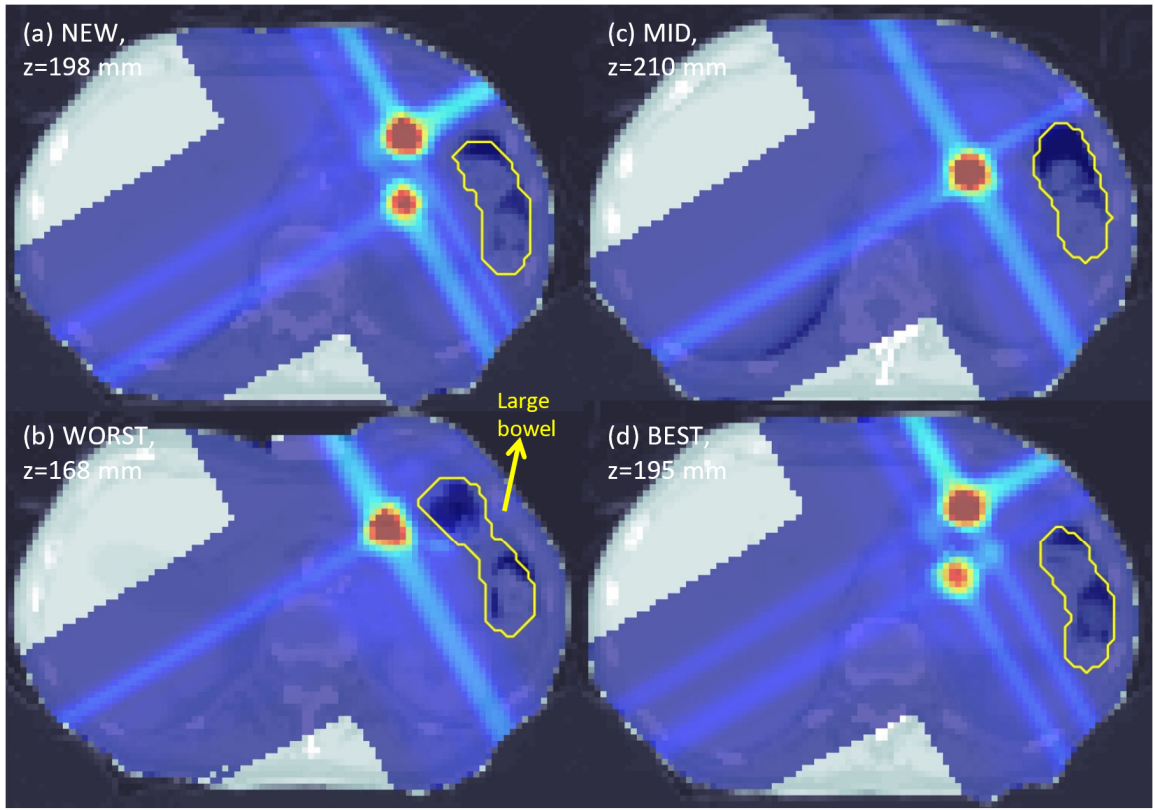

Figure 3. Axial dose slices for proton abdomen LATTICE plans using various techniques namely (a) NEW; (b) WORST; (c) MID; (d) BEST. Due to position differences of the lattice vertices between different techniques, the axial slices of the maximal PVDR from each plan are presented. The dose plot window is [0 Gy, 18 Gy].

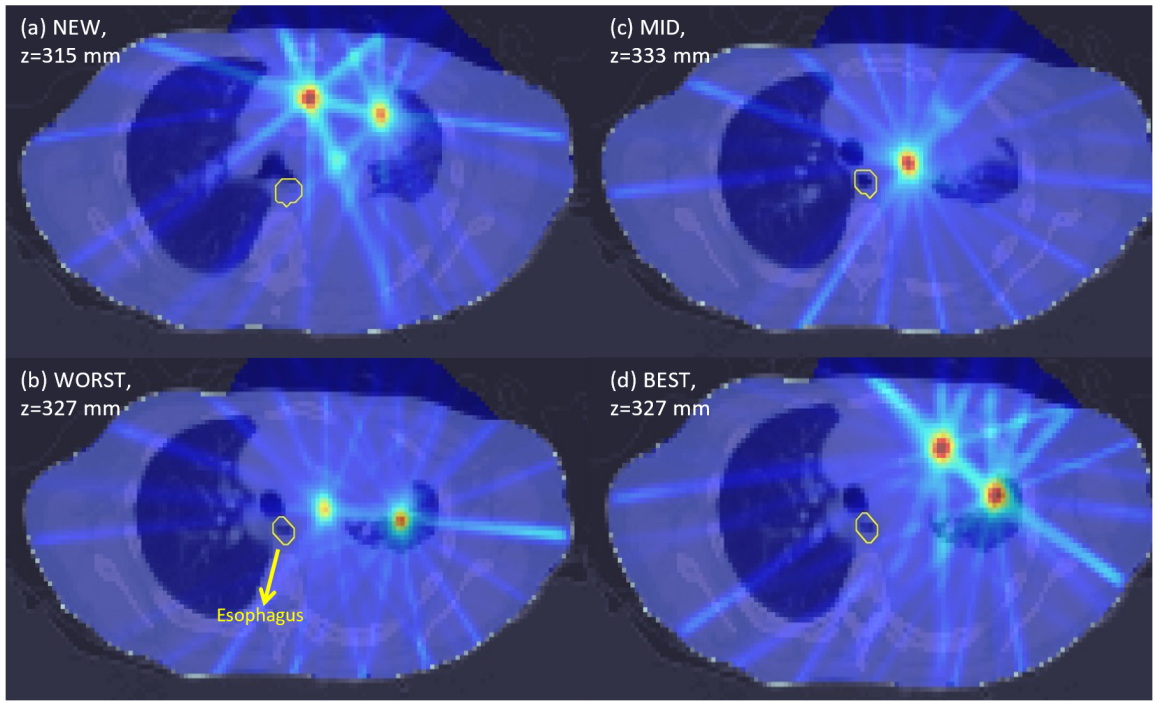

Figure 4. Axial dose slices for photon lung LATTICE plans using various techniques namely: (a) NEW; (b) WORST; (c) MID; (d) BEST. Due to position differences of the lattice vertices between different techniques, the axial slices at the midpoints of all lattice vertices from each plan are presented. Therefore, more peaks can be seen on the selected axial slices. The dose plot window is [0 Gy, 16 Gy].

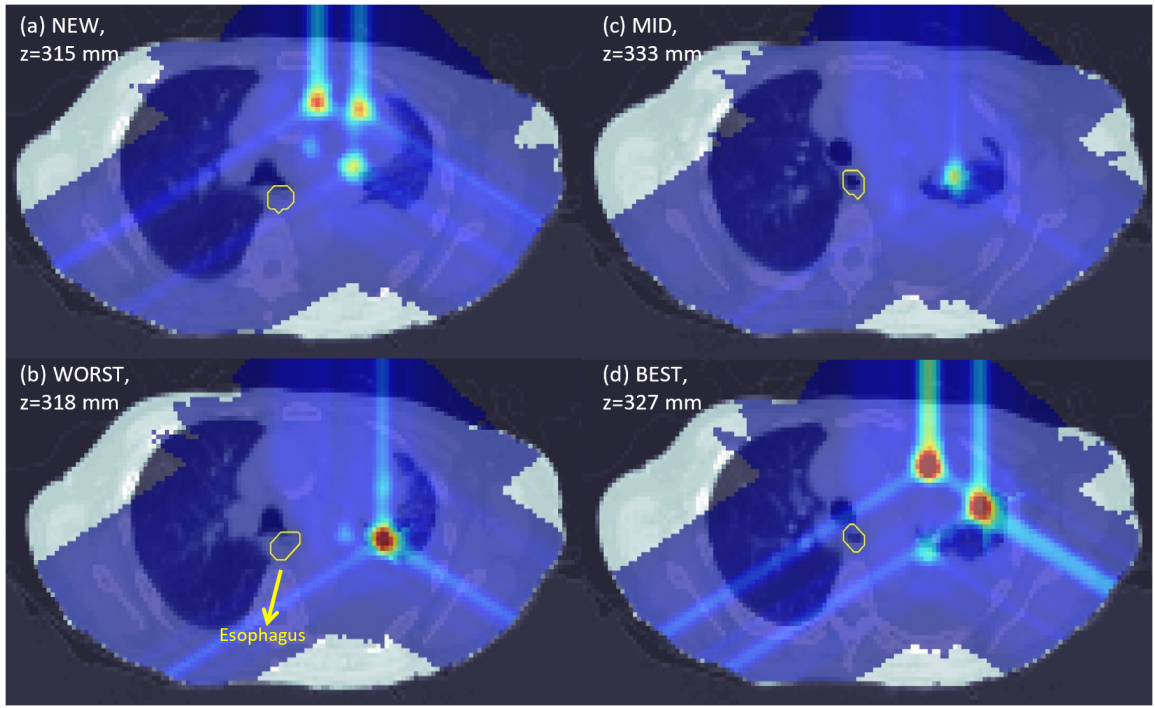

Figure 5. Axial dose slices for proton lung LATTICE plans using various techniques namely: (a) NEW; (b) WORST; (c) MID; (d) BEST. Due to position differences of the lattice vertices between different techniques, the axial slices at the midpoints of all lattice vertices from each plan are presented. Therefore, more peaks can be seen on the selected axial slices. The dose plot window is [0 Gy, 16 Gy].

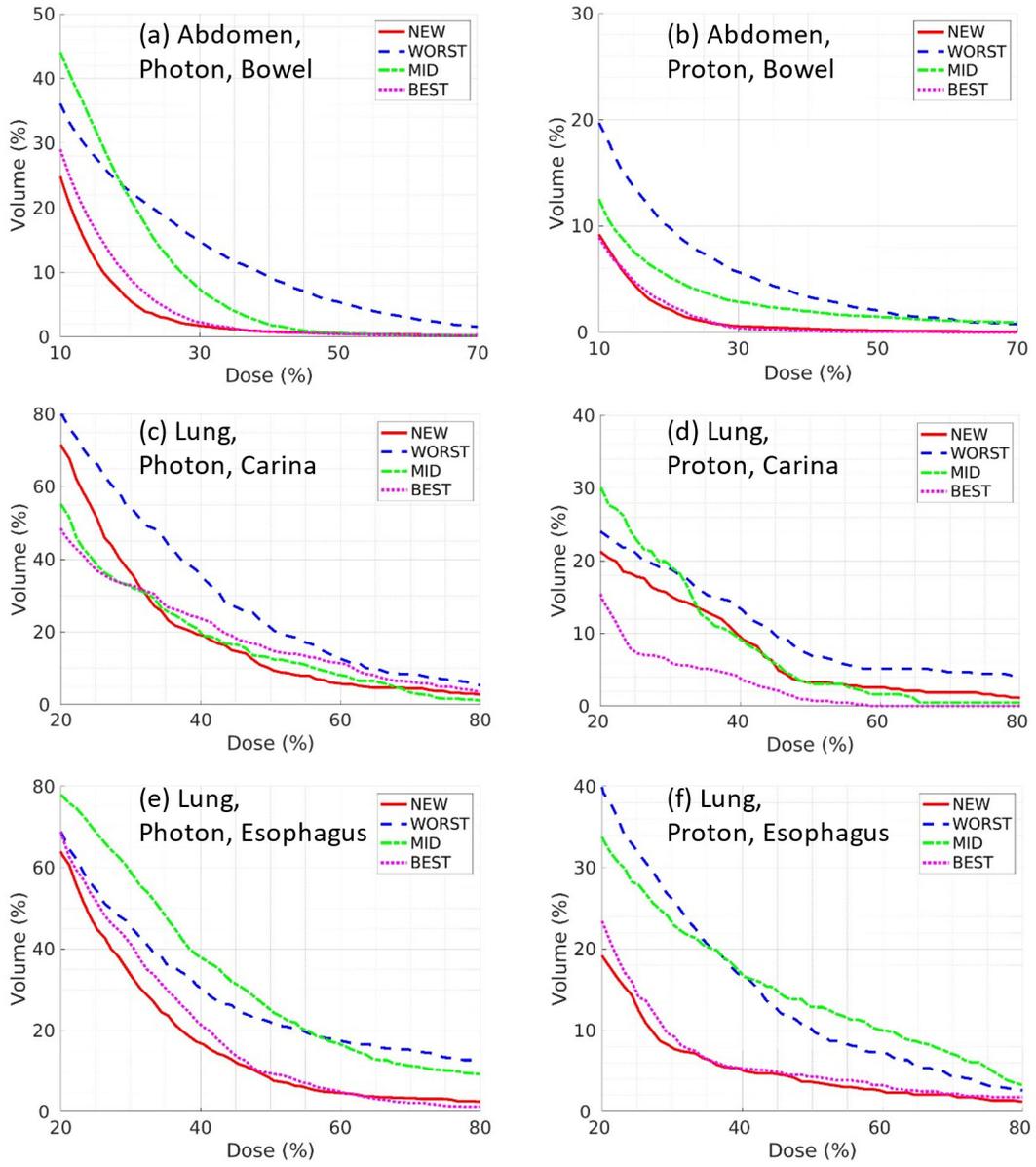

Figure 6. Comparisons of OAR DVH between different LATTICE techniques: (a) abdomen photon bowel; (b) abdomen proton bowel; (c) lung photon carina; (d) lung proton carina; (e) lung photon esophagus; (f) lung proton esophagus.

## 4. Discussion

In this work, we successfully developed a LATTICE treatment planning method that allows fully flexible high-dose vertex placement and enables the simultaneous optimization of vertex positions and dose automatically. The results showed that the NEW plans achieved comparable plan parameters (total objective value and PVDR) to the BEST plans, and significantly outperformed the MID and WORST plans. Our proposed method also provides a way to generate LATTICE plans with fully flexible vertex placement, which can provide additional options for patients with irregular tumor boundaries.

This new LATTICE treatment planning method simultaneously optimizes the fully flexible lattice positions and dose distribution by a constrained optimization problem in (5). Then, the problem was decoupled into subproblems by ADMM and solved separately. For the nonlinear subproblem (7a), the solutions of $(x^{k+1}, r^{k+1})$ were initially investigated using both the first-order gradient descent and second-order quasi-Newton methods and cross-compared. While the gradient descent method was found to decrease computational times due to reduced intermediate calculations, we realized that the second-order quasi-Newton method converges more stably to its minimum. Hence, all results were obtained using the second-order quasi-Newton method.

A potential limitation of our approach is the susceptibility of our methods to converge to a local minimum rather than a global minimum. This is because the newly formulated LATTICE optimization problem is nonconvex (e.g., the sigmoid function) and nonlinear (e.g., the joint optimization of $x$ and $r$). In practice, the solution convergence is sensitive to the choices of parameters, such as number of iterations, numerical step sizes, etc. Although the solutions can be theoretically better (i.e., smaller $F$ objectives) with more iterations and more informed parameter selections, we managed to achieve plan qualities (in terms of $F$) that are better than or comparable to BEST techniques with our NEW approach (e.g., Table 1), demonstrating its practical clinical utility. This is because BEST techniques do not consider fully flexible lattice vertex positioning and only sample a discrete set of initial lattice positions.

More parameters may be included as variables for optimization, such as peak dose, valley dose, lattice vertex size [31]. In addition, target PVDR and OAR dose can also be added to the objective to improve the plan quality. In this study, the proton beam angles were set empirically, while optimizing these angles could further reduce the dose to normal tissues. A joint beam angle and lattice position optimization method will be considered and can be solved via group-sparsity method [32].

## 5. Conclusion

A novel LATTICE treatment planning method has been introduced, providing fully flexible joint lattice position and dose optimization. This approach effectively optimizes target PVDR and OAR sparing, which is validated against plans from the conventional LATTICE method with regularly spaced and pre-selected lattice positions.